\documentclass[12pt]{article}
\usepackage[cp1251]{inputenc}
\usepackage[T1]{fontenc}
\usepackage{indentfirst}
\begin{document}

\begin{center}
{\Large Electro-Optical Radiation of a Charged Particle in a Single Laser Beam Field}
\end{center}

\bigskip

\begin{center}
{\large M.\,V.~Anokhin$^{1,2,3}$, V.\,I.~ Galkin$^{2}$, V.\,V. ~Kalegaev$^{2}$, G.\,A.~ Nagorsky$^{3}$,
V.\,I.~Panasiuk$^{2}$, and V.\,M.~Chabanov$^{3}$ } \\
\bigskip
{\small $^{1}$ Space Research Institute of the Russian Academy of Sciences, 84/32 Profsoyuznaya Str, Moscow, Russian Federation, 117997  \\
$^{2}$ Skobeltsyn Institute of Nuclear Physics, Federal State Budget Educational Institution of Higher Education M.V. Lomonosov Moscow State
University, Skobeltsyn Institute of Nuclear Physics (SINP MSU), 1(2), Leninskie gory, GSP-1, Moscow 119991, Russian Federation\\
$^{3}$ Institute of Natural High Technology, 31 Gorky Str, Tarusa, Kaluga Region,  249101, Russian Federation}
\end{center}

\bigskip

\begin{abstract}
A peculiar radiation arising as a result of radiation interference of nonlinear oscillators excited by a monochromatic
plane wave field of the incident particle is described. The radiation properties are determined by the fact that a phase
of each oscillator radiation fields is synchronized by a wave field, while the radiation itself occurs due to the particle field
influence on the oscillators. The consideration is performed for a thin film with negligible density effect. It is supposed that
the contribution is given only by a long-wave part of the Weizsacker spectrum for which nonlinear polarization coefficients
of medium are large.
\end{abstract}


\section{ Introduction}

In the classical work \cite{Ginzburg} it was noted that if the transition radiation occurs when a charged particle passes through the medium with
varying refraction coefficient one may talk of not only the transition radiation or the resonance transition radiation but call the corresponding
process a transition scattering. In fact, in that case a permittivity wave (refraction coefficient wave), which can be either standing or travelling,
as if scatters on the moving charge giving rise to an electromagnetic radiation. There will be no reason to apply for these conditions a term
{\it transition scattering} \cite{Ginzburg2} instead of transition radiation if the effect is not preserved in the limiting case of  stationary charge.
Then the discussion about the transition radiation is at any rate unnatural, whereas the term `transition scattering' is the essence of the matter.
In particular, this is in reference to, e.g., incidence of permeability wave on the stationary (fixed) charge, with e/m waves diverging from the charge.

At present, public opinion \cite{PlatFleish}  is in favor of the following determination of the transition radiation `{\it We call it the transition radiation, the part of
\textbf{multi-component medium} radiation that does not vanish in the limit of uniform source movement, and, vice versa, vanishes in the limit of homogeneous (at
any space and time scales) medium}'. This determination embraces both transition radiation on the specific boundary \cite{Ginzburg3}  and other types, such as
diffraction radiation \cite{BG}, resonance  radiation \cite{TerMik} and polarization  radiation \cite{Amusia}, etc.

The techniques of theoretical description of the radiation for different local dynamical multi-component systems have much in common. However,
taking into account  joint public opinion, in the present work we consider a dynamical system consisting of the non-linear optical medium in which a
monochromatic coherent e/m radiation scatters on a local region produced by the charged particle. It is shown that in the absence of density effect
there occurs a radiation which {\it can be called electro-optical}, because the Pockels effect is used. This phenomenon is practically non-inertial,
with speed capability of the order of $10^{-10}$ sec. Due to this fact, the electro-optical radiation can be used for determination of the Lorentz-factor of
charged particles in the large range (presumably, more than $10^{3}$), in particular, in experiments on the spacecraft instruments.

 \section{}
In this work a peculiar radiation is found that is brought about as a result of interference of non-linear film oscillator radiation,
when the oscillators are excited  by a monochromatic plane wave field of the incident particle.  The property of that radiation,
which we shall call an electro-optical radiation, is determined by the fact that the pase of  individual oscillator fields is synchronized
by a wave field, with the radiation itself induced by the particle field influence on the oscillators. It is supposed that
the contribution is given only by a long-wave part of the Weizsacker spectrum for which nonlinear polarization coefficients
of medium are large.

Let the radiation with the wavelength $\lambda$ be observed at the angle $\theta$ relative to the direction $\vec{e}_{3}$
of intersection between plane of the wave incidence and the film. Then interference condition has the form

\begin{eqnarray}
\cos(\theta)=\frac{\lambda}{d},
\label{cost}
\end{eqnarray}

\noindent where $d={\lambda}_{\kappa}/\cos(\psi)$, ${\lambda}_{\kappa}$ -- the wavelength of an external e/m field, $\psi$ is angle between the wave
propagation $\vec{e}_{\kappa}$ and the polar axis $\vec{e}_{3}$.

This condition, in view of linear extent of the sources, holds also in some neighborhood of generator of a cone with axis $\vec{e}_{3}$
and angular opening  $\theta$.  Consequently, as $\theta \rightarrow 0$ and $\lambda \rightarrow {\lambda}_{\kappa}/\cos(\psi)$  the radiation intensity grows.
So, the electro-optical radiation spectrum concentrates  near $\lambda = {\lambda}_{\kappa}/\cos(\psi)$ and, as will be shown below, its width and intensity
depend on energy of the incident particle.

In order to obtain the main formulas of the theory we shall use non-harmonic oscillator model, for which movement equation in the field $E^{(\gamma)}_{L}$
of the incident particle and laser field $E_{L}$ has the form

\begin{eqnarray}
\stackrel{\cdot \cdot}{X}_{L} + \Gamma_{f} \stackrel{\cdot}{X}_{L} + \omega^{2}_{f} X_{L} + T_{Lk}(x) X_{k}=
\frac{e_{f}}{m_{f}}(E_{L}+E^{(\gamma)}_{L}),
\label{nonhramonic}
\end{eqnarray}

\noindent where ${X}_{L}$ is an oscillator displacement from the equilibrium position $\vec{r}$, and  $\omega_{f}$, $\Gamma_{f}$, $e_{f}$ and
$m_{f}$ are frequency, attenuation, effective charge and mass of an oscillator, respectively; $T_{L k}(x)$ is the non-harmonic correction to frequency associated
with strong crystal fields in the film.

In case of the quadratic nonlinearity, this correction is particularly large when one of the fields is almost constant (analogous to the Pockels effect).

In the case under consideration a long-wave part of the Weizsacker spectrum of the particle is chosen as a slow field. The oscillator displacement $X_{L}$ takes
the form of a sum of slow $\rho_{L}$ and fast $\xi_{L}$ components. Neglecting from now on the terms, quadratic in $\xi_{L}$ and  $\rho_{L}$ and
corresponding to the radiation on the second and zeroth laser harmonics, we take, instead of (\ref{nonhramonic}), the following equation for $\xi_{L}$

\begin{eqnarray}
\stackrel{\cdot \cdot}{\xi}_{L} + \Gamma_{f} \stackrel{\cdot}{\xi}_{L} + \omega^{2}_{f} \xi_{L} + g(t) \sigma_{L k l} e^{(\gamma)}_{l}(\vec{\eta}) \xi_{k}=
\frac{e}{m} E_{L},
\label{nonhramonic2}
\end{eqnarray}

\noindent where $e^{(\gamma)}_{L}$ is direction of the transversal particle field component in the oscillator localization point $\vec{r}$, whereas a time
dependence of the frequency shift is given by

\begin{eqnarray}
g(t) l_{i}(\vec{r})=\int_{-\infty}^{\infty} \frac{E^{(\gamma)}_{L}(\omega, \vec{r})e^{i \omega t}}{\omega_{f}^{2}-i\omega \Gamma_{f}-\omega^{2}}d\omega.
\label{timedep}
\end{eqnarray}

Taking into account the correction to frequency as a perturbation, we get, in the first approximation, from Eq. (\ref{nonhramonic2}) for the Fourier transform
of  displacement $\xi_{L}$ the expression as follows

\begin{eqnarray}
\xi_{L}(\omega)=(\omega_{f}^{2}-i\omega \Gamma_{f}-\omega^{2})^{-1} \frac{e_{f}}{m_{f}} \{E_{L}(\vec{r},\omega) - \sigma_{i k l} e^{(\gamma)}_{l}(\vec{r})
\Pi_{k}(\vec{r},\omega)\},
\label{ksidispl}
\end{eqnarray}

\noindent where

\begin{eqnarray}
\Pi_{k}(\vec{r},\omega)=\int_{-\infty}^{\infty}\frac{E_{k}(\vec{r}, \nu)}{\omega_{f}^{2}-i \Gamma_{f}\nu-\nu^{2}}g(\omega-\nu)d\nu.
\label{piksi}
\end{eqnarray}

In the same approximation the solution to  the Maxwell equations for the space Fourier transform of the electro-optical radiation takes the form

\begin{eqnarray}
\vec{\stackrel{\sim}{E}}(\vec{k},\omega)=-\frac{\varepsilon(\omega)-1}{\varepsilon(\omega)} \,
\frac{\frac{\omega^{2}}{c^{2}}\varepsilon(\omega) \vec{\pi}(\vec{k})-\vec{k}(\vec{k}\vec{\pi}(\vec{k}))}{{\vec{k}}^{2}-\frac{\omega^{2}}{c^{2}} \varepsilon(\omega)},
\label{Etil}
\end{eqnarray}

\noindent where

\begin{eqnarray}
\varepsilon(\omega)=1+\omega^{2}_{p} \sum_{f}\frac{f}{\omega_{f}^{2}-i\omega \Gamma_{f}-\omega^{2}}.
\label{epsofom}
\end{eqnarray}

\noindent Here ${\omega}^{2}_{p}=4 \pi e^{2} N/m$ is plasma frequency,  $f$ denotes oscillator strength, $\vec{\pi}(\vec{k})$ -- space Fourier transform of the
vector (\ref{piksi}) summed over $f$, i.e., $\vec{\pi}(\vec{r})$ can be written as follows

\begin{eqnarray}
\pi_{L}(\vec{r})=\frac{\omega^{2}_{p}}{\varepsilon(\omega)-1} \, \frac{e}{m} \sum_{f} \frac{f\sigma_{ikl} e^{(r)}_{l}(\vec{r}) \Pi_{k}(\vec{r},\omega)}
{\omega_{f}^{2}-i\omega \Gamma_{f}-\omega^{2}}.
\label{pir}
\end{eqnarray}

Substituting Eqs. (\ref{timedep}) and  (\ref{piksi}) to (\ref{pir}) and carrying out integration over time we get for the vector $\pi_{L}(\vec{r})$  the following spectral representation

\begin{eqnarray}
\pi_{L}(\vec{r})=\frac{1}{\varepsilon(\omega)-1}\int_{-\infty}^{\infty}d\nu \varepsilon_{A}(\omega, \nu) \sigma_{Lkl} E^{(\gamma)}_{l}(\omega-\nu, \vec{r})
E_{k}(\nu, \vec{r}),
\label{pir2}
\end{eqnarray}

\noindent where $\varepsilon_{A}(\omega, \nu)$ is density of the linear electro-optical coefficient variance which is determined in the present model, for the tensor
$\sigma_{Lkl}$ independent of $f$, by the formula

\begin{eqnarray}
\varepsilon_{A}(\omega, \nu)=\frac{e}{m} \omega^{2}_{p} \sum f (\omega_{f}^{2}-i\omega \Gamma_{f}-\omega^{2})^{-1}
(\omega_{f}^{2}-i\nu \Gamma_{f}-\nu^{2})^{-1}  \nonumber \\
\times (\omega_{f}^{2}-i(\omega - \nu) \Gamma_{f}-(\omega - \nu)^{2})^{-1}.
\label{varepsA}
\end{eqnarray}

It is supposed, when evaluating the fields in the wave zone, that a distance to the observation point is large in comparison with the radiating system, but, on account
of good convergence of the integrals over the space occupied by the film, the size of the last can be taken infinite. The ordinary procedure of calculation gives for the
photon flux ${\cal N}(\omega)$ the following expression

\begin{eqnarray}
\frac{d{\cal N}(\omega)}{dO d\omega}=\frac{\omega^{3}n_{1}}{2 \hbar c^{3}}(\delta_{pq}-n_{p}n_{q})\sigma_{pjk}\sigma_{qj'k'} \int \int
\frac{d^{3}\vec{r}\,' d^{3}\vec{r}\,''}{(4 \pi ^{2})} \nonumber \\
\times \exp\{i \frac{\omega}{c} n_{1} \vec{n}(\vec{r}\,''-\vec{r}\,')\} \cdot \int_{-\infty}^{\infty} \int_{-\infty}^{\infty} d\nu_{1} d\nu_{2} \varepsilon_{A}(\omega, \nu_{1})
\varepsilon_{A}(-\omega, \nu_2)  \nonumber \\
\times E^{(\gamma)}_{k}(\omega-\nu_{1}, \vec{r}\,') E^{(\gamma)}_{k'}(-\omega-\nu_{2}, \vec{r}\,'') E_{j}(\nu_{1}, \vec{r}\,') E_{j'}(\nu_{2}, \vec{r}\,''),
\label{photnflx}
\end{eqnarray}

\noindent where $\vec{n}$ denotes direction to the observation point, $n_{1}=Re \sqrt{\varepsilon(\omega)}$.

Let us confine ourselves in further consideration to the case when the laser field can be represented in the form of a finite superposition of plane waves

\begin{eqnarray}
E_{\nu}(t, \vec{r})=\sum_{\kappa} E^{(\kappa)}_{\nu} \cos\{\omega_{\kappa}(\frac{n_{\kappa}}{c} \vec{e}_{\kappa} \vec{r} - t) +\psi_{\kappa}\},
\label{Eplwvs}
\end{eqnarray}

\noindent where $E^{(\kappa)}_{j}$, $\psi_{\kappa}$ are amplitude and phase shift of the wave, respectively, n is refraction coefficient of the film on the frequency
$\omega_{\kappa}$, and, as the particle field $E^{(\kappa)}_{L}=e^{(\kappa)}_{L}(\vec{r})E^{(\gamma)}$, we substitute the Fourier transform $E^{(\gamma)}(\omega, \vec{r})$
of the transversal component of the field generated by a uniformly moving charge \cite{Jaks}

\begin{eqnarray}
E^{(\gamma)}(\omega, \vec{r})=\frac{e}{\pi \beta^{2}}  \frac{\omega e^{i \omega t_{0}}}{c^{2} \varepsilon_{0}(\omega)}\sqrt{1-\beta^{2}\varepsilon_{0}(\omega)} K_{1}
\frac{\omega b}{\beta c} \sqrt{1-\beta^{2} \varepsilon_{0}(\omega)},
\label{Eunmch}
\end{eqnarray}

\noindent where $\varepsilon_{0}(\omega)$ is equal to unity in the absence of density effect (a medium polarization induced by particles with $\beta \to 1$), $b$ is an	impact parameter,
determined by the formula

\begin{eqnarray}
\vec{b}=\vec{r}-\vec{r}_{0} - \vec{s}(\vec{s}\vec{r}-\vec{s}\vec{r}_{0}), \,\,  e^{(\gamma)}_{L}(\vec{r})=\frac{b_{L}}{b}.
\label{b}
\end{eqnarray}

\noindent Here $\vec{s}$ is direction of the particle movement,  $\vec{r}_{0}$ $t_{0}$ are the coordinates of incident point and the moment of incidence time.

Taking into account Eqs. (\ref{Eplwvs}), (\ref{Eunmch}) and (\ref{b}) the integration of (\ref{photnflx}) is elementary, but we do not give a general formula for a plate of arbitrary
width $\Delta$ because the radiation intensity dependence being of interest to us, which is a function of energy particle, is possible only for thin films for which the density effect is missing.

Let us assume further on that the following condition holds

\begin{eqnarray}
\Delta < < \frac{2\pi c}{\omega_{\kappa}}
\label{muchl}
\end{eqnarray}

\noindent and omit the terms corresponding to contributions of the high-frequency Fourier components of the particle field for which phenomenological constants
$\varepsilon_{A}$ have to be small in a more realistic model than ours. This allows us to represent the result of integration  (\ref{photnflx}) in the form

\begin{eqnarray}
\frac{d{\cal N}(\omega)}{dO d\omega}=\frac{e^{2}\omega^{3}n_{1}\Delta^{2}}{32 \pi^2 \hbar c^{5} \beta^{2}}(\delta_{pp'}-\Pi_{p}\Pi_{p'})\sigma_{pjL}
\sigma_{p'j'L'} \sum_{\kappa,\kappa'} E_{j}^{(\kappa)} E_{j'}^{(\kappa')} \nonumber \\
\times Re\{\exp[i(\phi^{(\kappa)}_{0}-\phi^{(\kappa')}_{0})] \varepsilon^{*}_{A}(\omega, \omega_{\kappa})\varepsilon_{A}(\omega, \omega_{\kappa '})\}
\lambda_{L}\lambda{'}_{L'} D^{-1}(\vec{\lambda}, \kappa)  D^{-1}(\vec{\lambda}', \kappa'),
\label{photnflx2}
\end{eqnarray}

\noindent where phase shifts  $\phi^{(\kappa)}_{0}$ depend on the position and  time moment of the incoming particle incidence on the film as follows

\begin{eqnarray}
\phi^{(\kappa)}_{0}=n_{\kappa} \frac{\omega_{\kappa}}{c} \vec{e}_{\kappa}\vec{r}_{0} - \omega_{\kappa} t_{0} + \psi_{\kappa}.
\label{phasesh}
\end{eqnarray}

\noindent In the Eq. (\ref{photnflx2}) the interference denominators  $D(\vec{\lambda}, \kappa)$ have the form

\begin{eqnarray}
D(\vec{\lambda}, \kappa)=\vec{\lambda}^{2}+(\frac{\omega-\omega_{\kappa}}{\omega n_{1} \beta \gamma}S_{1})^{2}
\label{interfden}
\end{eqnarray}

\noindent and vector $\vec{\lambda}$ is determined by the relation

\begin{eqnarray}
\vec{\lambda}=\vec{s} \times (\vec{e} \times (\frac{\omega}{c}n_{1}\vec{n}-\frac{\omega_{\kappa}}{c} n_{\kappa} \vec{e}_{\kappa})).
\label{lambda}
\end{eqnarray}

Everywhere $\vec{e}$ denotes a unit normal vector to the surface of the film.

It is worth to mention some peculiarities of differential flux of the emitted photons (\ref{photnflx2}).

If the film is illuminated by more than one laser the flux, and hence a full number of photons, depends on the position and time moment of the particle
incidence on the film. In principle, this makes it possible to evaluate the particle coordinates  with  accuracy of
the order of $c/(\omega_{\kappa} - \omega_{\kappa'})$   and record the charge passage  through the film with time accuracy $\sim 1/(\omega_{\kappa} - \omega_{\kappa'})$.

Second peculiarity of the electro-optical radiation is associated, as was already mentioned, with its dependence on the particle energy. We consider now
the character of that dependence  using evaluation of one-laser wave photon spectrum as an example.

Let the radiation direction $\vec{n}$ be specified by polar angles $\{\theta, \varphi\}$ relative to the unit vector system $\{\vec{e}_{1}, \vec{e}_{2}, \vec{e}_{3}\}$ as follows

\begin{eqnarray}
\vec{n}=\vec{e}_{3} \cos \theta +(\vec{e} \cos \varphi +\vec{e}_{2} \sin \varphi ) \sin \theta ,
\label{raddir}
\end{eqnarray}

\noindent where $ \vec{e}_{2}= \vec{e}_{3} \times  \vec{e}$. Then the vector $\vec{\lambda}$ in accordance with (\ref{lambda}) is determined by the formula

\begin{eqnarray}
\vec{\lambda}=\frac{\omega n_{1}}{c} \vec{s}  (q \vec{e}_{2} + \sin \theta \sin \varphi \vec{e}_{3}),
\label{lambda2}
\end{eqnarray}

\noindent where

\begin{eqnarray}
q=\frac{\omega_{\kappa} n_{\kappa}}{\omega n_{1}}\cos \psi - \cos \theta,
\label{q}
\end{eqnarray}

\noindent $\psi$ being the angle between the direction of wave propagation and the axis $\vec{e}_{3}$, i.e., $\cos \psi=\vec{e}_{3} \vec{e}_{\kappa}$.

On the assumption that the difference of frequencies $\omega-\omega_{\kappa}$ is of the order of lattice vibration eigenfrequencies in a crystal, we use
the connection between the tensor $\sigma _{Ljk}$ and the linear electro-optical coefficient $r_{Ljk}$ (the Pockels effect) which corresponds to the case
when arguments of the function $\varepsilon_{A}(\omega,\omega ')$ coincide, i.e.,

\begin{eqnarray}
r_{Ljk}=\sigma_{Ljk} \varepsilon ^{-2}(\omega) \varepsilon_{A}(\omega, \omega)
\label{rlj}
\end{eqnarray}

\noindent in order to get the photon spectrum gauge provided that $Im \varepsilon_{A}(\omega_{1},\omega)=0$ and the absorbtion is zero.

Allowing for the fact that  the product $r_{Ljk} E_{j}=R_{Lk}$ is dimensionless, we obtain from (\ref{photnflx2}) after integration over $\varphi$ a spectrum of the form

\begin{eqnarray}
\frac{d{\cal N} \omega}{d \omega}=\frac{e^{2} \omega \Delta ^{2}\varepsilon ^{4}(\omega)}{8\pi \hbar c^{3} \beta ^{2} n_{1}} G,
\label{photnflx3}
\end{eqnarray}

\noindent where the integral

\begin{eqnarray}
G=R_{pL}R_{p'L'} \int \frac{d O}{4\pi}(\delta_{pp'}-n_{p}n_{p'}) \frac{\lambda_{L}\lambda_{L'}}{D^{2}(\lambda, \kappa)}(\frac{\omega n_{1}}{c})^{2}
\label{G}
\end{eqnarray}

\noindent can be rewritten, after summing over the indexes using the following notations

\begin{equation}
\begin{array}{lcr}
A_{1}=S_{1}R_{12}-S_{2}R_{11} & , & B_{1}=S_{1}R_{13}-S_{3}R_{11} \\
A_{2}=S_{1}R_{22}-S_{2}R_{12} & , &  B_{2}=S_{1}R_{23}-S_{3}R_{12} \\
A_{3}=S_{1}R_{23}-S_{2}R_{13} & , & B_{3}=S_{1}R_{33}-S_{3}R_{13},
\end{array}
\label{notats}
\end{equation}

\noindent in the form

\begin{eqnarray}
G=\frac{A_{1}^{2}-A_{2}^{2}}{(1-S^{2}_{3})^{2}}+\frac{1}{2\sqrt{2}} \int_{0}^{\pi} \frac{\sin \theta d\theta}{(az)^{1/2}}\{g_{1} q^{2}(a_{+}^{1/2}+a_{-}^{1/2}) \nonumber \\
\times (\frac{2 \stackrel{\sim}{\epsilon}-z}{a}+\frac{\stackrel{\sim}{\epsilon}+a^{1/2}}{a^{1/2} z})+g_{e} \sin ^{2}\theta (a_{+}^{1/2}+a_{-}^{1/2})
(\frac{2 \stackrel{\sim}{\epsilon}-z}{a}-\frac{\stackrel{\sim}{\epsilon}}{a^{1/2} z}) \nonumber \\
- g_{3} \frac{8S_{2}S_{3}q^{2}\sin ^{2}\theta}{a_{+}^{1/2}+a_{-}^{1/2}} (\frac{2 \stackrel{\sim}{\epsilon}-z+a^{1/2}}{a}+\frac{\stackrel{\sim}{\epsilon}+a^{1/2}}{za^{1/2}}) \nonumber \\
- g_{4} \frac{8S_{2}S_{3}q \sin ^{4}\theta}{a_{+}^{1/2}+a_{-}^{1/2}} (\frac{2 \stackrel{\sim}{\epsilon}-z+a^{1/2}}{a}+
\frac{\stackrel{\sim}{\epsilon}}{za^{1/2}})+(A_{1}^{2}-A_{2}^{2}) \nonumber \\
\times \frac{a_{+}^{1/2}+a_{-}^{1/2}}{(1-S_{3}^{2})^{2}}[(1-S_{3}^{2}) \sin ^{2}\theta (\stackrel{\sim}{\epsilon}-z)
(\frac{4 \stackrel{\sim}{\epsilon}-2z}{a}+\frac{\stackrel{\sim}{\epsilon}}{za^{1/2}})-(\stackrel{\sim}{\epsilon}-z)^{2} \nonumber \\
\times \frac{2 \stackrel{\sim}{\epsilon}-z}{a} + 3 \stackrel{\sim}{\epsilon} -2z]\}.
\label{G2}
\end{eqnarray}

\noindent Here the following notations are used

\begin{equation}
\begin{array}{l}
\stackrel{\sim}{\epsilon}=\epsilon+(1-S^{2}_{2})q^{2},  \quad \epsilon=(\frac{\omega-\omega_{\kappa}}{\omega n_{1}\beta}S_{1})^{2}(1-\beta ^{2}),  \\
a=a_{+}a_{-}, \quad  a_{\pm}= \stackrel{\sim}{\epsilon} \pm 2S_{2}S_{3} q \sin \theta + (1-S^{2}_{3}) \sin ^{2} \theta \\
z=2 \stackrel{\sim}{\epsilon} -\frac{1}{2}(a^{1/2}_{+}-a^{1/2}_{-})^{2},  \quad \vec{S}=S_{1}\vec{e}+S_{2}\vec{e}_{2}+S_{3}\vec{e}_{3},
\end{array}
\label{notats2}
\end{equation}

\noindent and

\begin{equation}
\begin{array}{l}
g_{1}=B^{2}_{1} \cos ^{2}\theta + b^{2}_{2} + B^{2}_{3} \sin ^{2}\theta, \\
g_{2}=A^{2}_{1} \cos^{2}\theta + A^{2}_{2} + A^{2}_{3} \sin ^{2}\theta + 2(A_{2}B_{3}+A_{3}B_{2})q \cos \theta + (B^{2}_{1}-B_{2}^{2})q^{2}, \\
g_{3}=A_{1}B_{1} \cos ^{2}\theta  + A_{2}B_{2} +A_{3}B_{3}\sin ^{2}\theta + B_{2}B_{3}q\cos \theta , \\
g_{4}=A_{2}A_{3} \cos \theta + (A_{1}B_{1}-A_{2}B_{2})q.
\end{array}
\label{notats3}
\end{equation}

The terms in (\ref{G2}) growing with energy correspond to the integrals diverging when $\epsilon \to \infty$. Selecting these terms we get the asymptotic form
$G \rightarrow G^{(\infty)}$  as follows

\begin{eqnarray}
G^{(\infty)}=\frac{1}{2\sqrt{2}} \int_{0}^{\pi} \frac{\sin \theta d\theta}{(a)^{1/2} z^{3/2}}\{(g_{1} q^{2} + g_{2}\frac{\stackrel{\sim}{\epsilon}\sin ^{2} \theta}{a^{1/2}})
(a_{+}^{1/2}+a_{-}^{1/2}) \nonumber \\
- 2g_{3}q \sin \theta (a_{+}^{1/2}-a_{-}^{1/2})\}.
\label{Gasymp}
\end{eqnarray}

What is more, only three first terms are left in (\ref{notats3}) and free of $q$. In fact, since the integration (\ref{G2}) is performed for finite limits,  denominator vanishing is needed
for the divergence, i.e., the following condition must be satisfied

\begin{eqnarray}
z=\sqrt{\hat{\epsilon}\, ^{2} +4[\epsilon (1-S_{3}^{2}) + S_{1}^{2} q^{2}] \sin ^{2}\theta} + \hat{\epsilon} \rightarrow 0,
\label{ztozero}
\end{eqnarray}

\noindent where

\begin{eqnarray}
\hat{\epsilon}= \epsilon + (1-S_{2}^{2})  - (1-S_{3}^{2}) \sin ^{2}\theta
\label{epsil}
\end{eqnarray}

\noindent and the expression under root sign is equal to $a$. This condition is valid if and only if  $q \to 0$ together with $S$. It is possible that $a \ne 0$, i.e., the angle
$\theta$ remains finite. That means that the radiation frequency is fixed by the condition (\ref{q}) with $q=0$, which coincides with the interference condition (\ref{cost}).

When $\omega>\frac{n_{\kappa}}{n_{1}}\omega_{\kappa}\cos \psi$ Eq. (\ref{Gasymp}) is readily integrated and the radiation spectrum has the form (\ref{photnflx3}) with
$G=G^{(\infty)}_{+}$ where

\begin{eqnarray}
G^{(\infty)}_{+}=\frac{1-S_{3}^{2}}{2\sqrt{2}S^{3}_{1}}[1-(\frac{\omega_{\kappa}n_{\kappa}}{\omega n_{1}} \cos \psi)^{2}]^{-1/2}
(\frac{S_{1}^{2}g_{2}}{(1-S_{3}^{2})^{2}}+\bar{g}_{1})\ln\frac{4 S^{2}_{1}}{\epsilon(1-S_{3}^{2})},
\label{Ginfty}
\end{eqnarray}

\noindent where $\bar{g}_{1}$ is obtained from $g_{1}$ by substitution $B_{L}\rightarrow B_{L}-\frac{S_{2} S_{3}}{1-S_{3}^{2}}A_{L}, \, L=1,2,3$ and the angle $\theta$
is fixed by the condition $q=0$. Please note logarithmic growth of this part of the spectrum when incident particle energy increases and singularity at $\omega=\omega_{\kappa}$
because of $\epsilon=0$ on frequency of the incident wave.

In the domain $\omega<\frac{n_{\kappa}}{n_{1}} \omega_{\kappa} \cos \psi$  radiation in the limit $\epsilon \rightarrow 0$ is independent of energy because $q \neq 0$ on entire interval
of the angles and all the integrals are limited. As a matter of fact, the radiation in this domain is a background one because here we need to allow for the interference with other radiations
(Cherenkov and transition radiations, nonlinear incoherent light scattering and so on).

Finally, in region of the frequencies asymptotically close to a base frequency of the electro-optical radiation

\begin{eqnarray}
\omega \sim \frac{n_{\kappa}}{n_{1}} \omega_{\kappa} \cos \psi
\label{omsim}
\end{eqnarray}

\noindent the spectrum demonstrates the power-like dependence on energy. This region is characterized by an additional singularity in the integrand of Eq. (\ref{Gasymp}) as $a \to 0$.

Let $\omega =\frac{n_{\kappa}}{n_{1}} \omega_{\kappa} \cos \psi $, then $q=1-\cos \theta$ and when $q \to 0$ we have $\theta \to 0$. Evaluating a basic term in Eq. (\ref{Gasymp})
gives the spectral density at this point of the form (\ref{photnflx3}) with $G=G^{(\infty)}_{0}$, where

\begin{eqnarray}
G^{(\infty)}_{0}=\frac{\Gamma^{2}(1/4)}{16 \pi ^{1/2}}\frac{(1-S_{3}^{2})^{3/4}}{S^{S/2}_{1} \epsilon^{1/4}} [\frac{2S^{2}_{1} g_{2}}{(1-S_{3}^{2})^{2}} + \bar{g}_{1}].
\label{Ginfty0}
\end{eqnarray}

The coefficient $\epsilon^{-1/4}$ corresponds to the spectrum growth  proportional to the energy to the power 1/2.

Clarification of details of the spectrum near the the asymptotic peak (\ref{omsim}) is unpractical in this example because the form of basic line of the electro-optic radiation essentially depends
on physical conditions of the interference such as number of plane waves, the film configuration, spectral density of the particle field (for example, when the external magnetic field is applied this quantity is significantly changed), nonlinear corrections to the density effect, resonance effects and so on. Ascertaining the influence of many of these conditions is beyond the scope of our simple
model of inharmonic oscillators.

\end{document}